# Efficient distribution and improved security for reliable cloud storage system

Ninoslav Marina, Aneta Velkoska, and Natasha Paunkoska [*][†][‡]

February 5, 2016

**Abstract**

The distributed data storage systems are constructed by large number of nodes which are interconnected over a network. Each node in such peer-to-peer network is vulnerable and at a potential risk for attack. The attackers can eavesdrop the nodes and possibly modify their data. Hence distributed storage systems should be secure apart from satisfying the reconstruction and repair requirements. We constructed a distributed storage system, Twin MDS code framework which is more efficient than the regenerating codes based storage systems. We prove that this Twin MDS code framework gives better performance than MBR codes and equal with MSR codes in the distribution process and investigate its security performance comparing with the security of the MBR and MSR codes. Such Twin MDS code framework is examined in an eavesdropper model where passive attackers can access to the stored data or/and downloaded data during the repair process. We demonstrate that the Twin MDS code framework manages better results than MBR and MSR codes regarding the security in the system.

## 1 Introduction

Cloud storage is often implemented by complex multitiered distributed systems on clusters of thousands of commodity servers. These systems known as distributed storage systems (DSS) store different types of data-files (messages) dispersed across the distributed servers (nodes) in the network. DSS are commonplace nowadays, they operate in several environments such as peer-to-peer

[*]N. Marina and A. Velkoska are with the Faculty of Communication Networks and Security, University of Information Science and Technology, Ohrid, Macedonia.

[†]N. Paunkoska is with Faculty of Information Systems, Visualization, Multimedia and Animation, University of Information Science and Technology, Ohrid, Macedonia.

[‡]

(P2P) systems and data centers that comprise the backbone infrastructure of cloud computing. Main advantage offered by the distributed storage systems is reliable and cost-effective storage of large amounts of data. With the number of its components increasing (storage nodes, but also routers, network, power supply, cooling, etc.), an DSS ends up having a significant (even if small) subset of these components not functioning properly at almost any time instance. Thus, fault tolerance to make the overall system and its services transparent from the underlying faults is essential. This is achieved by the addition of redundancy.

Redundancy is accomplished in many different manners.The most simplest and common form applied in these systems for achieving the redundancy is by using replication schemes. Three-times data replication as an industrial standard is the simplest form of protective data redundancy, with two more additional copies of the original object being created and maintained to be available if the original gets lost. The more redundancy is used, the more fault-tolerant the DSS becomes. But the redundancy increases the overheads of the storage infrastructure. Data to be stored is not reducing overtime, and a study sponsored by the information storage company EMC estimated that the worlds data is more than doubling every two years. [1]

The demand for data storage is astonishing, due to the new emerging applications like social networks analysis, semantic Web analysis, bioinformatics network analysis, city sensing and monitoring, and additional variety of data that should be managed daily through the communication broadband networks. For instance, human beings now create 2.5 PB bytes of data per day. The rate of data creation has increased so much that 90% of the data in the world today has been created in the last two years alone [2]. Significant impact from the emerging world also have big companies, governments, militaries, banks and any kind of organization and group that have need for storage of big amount of sensitive information.

Towards finding optimal solution, some of the challenges for managing the enormous amount of data and efficient processing, storage and maintaining of the data are elaborated in [3–6]. A smarter solution with less redundancy and better reliability is offered by erasure codes. Maximum Distance Separable (MDS) codes are erasure codes that are good choice for successful reconstruction of the entire message that is performed by the user (data collector).

Let $S$ is the size in symbols (can be bits or digits in general) of the data file that needs to be stored in the distributed network. Suppose that this file of size $S$ is divided into $k$ pieces which are mapped to $n$ encoded fragments using an $(n, k)$ MDS code, and all encoded pieces are stored on $n$ distinct nodes in the network. The data can be retrieved by accessing the encoded fragments from any $k$ of these $n$ nodes, and carrying out a reconstruction (decoding). When a node fails, $\alpha = \frac{S}{k}$ symbols are affected. Therefore, a new replacement node (newcomer) must be added in the network that will perform the repair process for the missing data. This process allows the newcomer to contact at least $k$ active nodes, downloads the entire data stored on them and after that to execute adequate operations for extracting the exact lost data stored on the failed node.

This strategy is a waste of communication bandwidth if only one encoded fragment is needed, though its cost is depreciated if repairs are delayed, and multiple repairs are carried out together. Even in the case of delayed repair, the newcomer remains a bottleneck, which might slower the repairs. Meaning, a slower repair process can in turn adversely affect the long term data durability. Therefore, a new concept of codes called regenerating codes which can be seen as a combination of an erasure code and a network code initially was proposed by Dimakis et al. [7].

In the regenerating codes the parameter $d$ is the number of nodes out of $n-1$ that are going to be contacted during the repair process and the parameter $\beta$ ($\beta \leq \alpha$) is the size of data that will be downloaded from each of the $d$ nodes. The correlation between the total downloaded amount $d\beta = \gamma$ is known as repair bandwidth and the storage $\alpha$ is studied in [8–11] for achieving better results.

The parameters in the regenerating code that aim to store the maximum file size $S$ reliably must satisfy the following condition examined in [7]

$$S \leq \sum_{i=0}^{k-1} \min\left\{\alpha, (d-i)\,\beta\right\}. \tag{1}$$

Based on the tradeoff between $\alpha$ and $\gamma$ the two extreme points can be obtained presented in [7]. In the reconstruction case when the storage per node $\alpha$ tends to be at least $\frac{S}{k}$ the extreme point is termed as the Minimum Storage Regeneration (MSR) point. The MSR point is achieved by the pair

$$(\alpha_{MSR}, \gamma_{MSR}) = \left(\frac{S}{k}, \frac{S}{k}\frac{d}{d-k+1}\right). \tag{2}$$

Otherwise, in a case when the repair bandwidth $d\beta$ is equal to $\alpha$, the extreme point is referred as the Minimum Bandwidth Regeneration (MBR) point that is achieved by the pair

$$(\alpha_{MBR}, \gamma_{MBR}) = \left(\frac{S}{k}\frac{2d}{2d-k+1}, \frac{S}{k}\frac{2d}{2d-k+1}\right). \tag{3}$$

Since both extreme conditions cannot be satisfied at the same time, K . V. Rashmi et al. [12] propose a new concept for distributed storage network, called Twin–code framework that eases data-reconstruction and node-repair during failure of some nodes in the network.

Because the distributed data storage system is formed by many nodes widely spread across the Internet, each node in such peer-to-peer network is vulnerable and a potential point for attack. The attackers can eavesdrop the nodes and possibly modify their data. Securing such data from adversaries/eavesdroppers is necessary to ensure data secrecy for the users. Hence DSS should be secure apart from satisfying the reconstruction and repair requirements. Some researches that covers the topic of security are formulated in [13–20]. This paper, besides the reliability of the stored data addresses the issue of its security. In our analysis we concentrate on a passive eavesdropper who can eavesdrop on nodes in the system or/and on newcomers during the repair process.

The rest of the paper is organized as follows: Section II recalls the Twin-code framework functioning and its advantages. In Section III by comparing the reconstruction and repair process of minimum bandwidth (MBR) and minimum storage regenerating (MSR) codes vs. the Twin-codes, we show that the number of message symbols that can be distributed in the Twin-code storage system is greater than the number of message symbols distributed by MBR codes and equal with the number of message symbols distributed by MSR codes. In Section IV we construct a new secure Twin-code framework in presence of passive adversary and show that our proposed secure scheme gives better secrecy performance than both MBR and MSR codes. Section V concludes the paper.

## 2 Twin-code framework

In the Twin–code framework all nodes are partitioned into two groups: nodes of Type 1 (circles) and nodes of Type 2 (squares) as depicted in Fig.1. To achieve the data stored on Type 1 nodes, the message is encoded using a linear code $\mathcal{C}_1$; and the data on Type 2 nodes, first the message symbols are permuted by transposition and then encoded by a second linear code $\mathcal{C}_2$. It is not necessary the two codes to be distinct. In the case of data reconstruction, the data collector will contact a feasible subset of nodes of the same type for recovering the entire message, as shown in Fig. 1 (a).

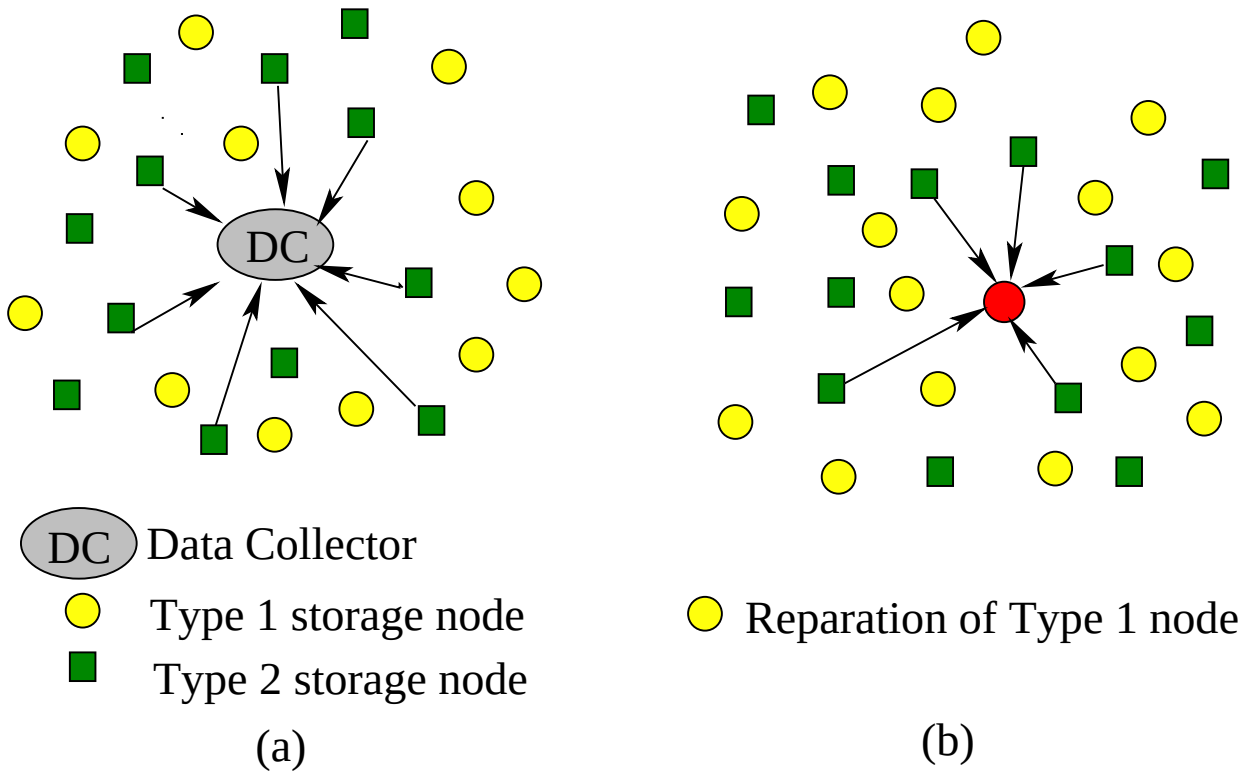

Figure 1: Twin-code framework: (a) Data reconstruction. (b) Repair process.

When a node from each type fails, the repair process is accomplished by downloading data from a feasible subset of nodes from the opposite type, as shown in Fig. 1 (b).

Without loss of generality, we can separate the nodes in the Twin-code framework into systematic and parity nodes. The node is systematic if the stored symbols on it are original. And if the symbols on the node are not pure original, but some combination of them, then the node is parity. We assume repair process of only systematic nodes.

## 2.1 Encoding

There are $n_1$ Type 1 nodes, and $n_2$ Type 2, where $n = n_1 + n_2$ is the total number of storage nodes in the framework. Note that the storage nodes of both types have the same characteristics.

In the sequel we use the following notations:

1. The original message consists of $S_{\mathrm{TW}}$ symbols which belong to a finite field $\mathbb{F}_q$;
2. $\mathcal{C}_i$, $i = 1, 2$ is an arbitrary $(n_i, k)$ linear code over $\mathbb{F}_q$ with generator matrix $G_i$;
3. $\mathbf{g}_{(i,j)}$ for all $1 \leq j \leq n_i$ is the $j$-th column of $G_i,\ i = 1, 2$.

The original message is first split into $k$ fragments of $k$ symbols, such that, $S_{\mathrm{TW}} = k^2$. Hence, these $S$ symbols are arranged into a square $(k \times k)$ matrix $A_1$. This matrix is called a message matrix. Let

$$A_2 \triangleq A_1^T, \tag{4}$$

where the superscript $T$ denotes a transpose of a matrix. For $i = 1, 2$ each node of Type $i$ stores $k$ symbols from the appropriate column of the $(k \times n_i)$ matrix $A_i G_i$, i.e., in the node $j$ $(1 \leq j \leq n_i)$ of Type $i$ we store the symbols from the $j$-th column of the matrix $A_i G_i,\ i = 1, 2$ in Fig. 2 defined by

$$A_i \mathbf{g}_{(i,j)}. \tag{5}$$

With this encoding algorithm every node stores $k$ symbols and each node $j$ of Type $i$ is associated with a different column $\mathbf{g}_{(i,j)}$ of $G_i$, called encoding vector of that node. With this algorithm the data is encoded and mapped into the network.

## 2.2 Twin MDS Codes for data – reconstruction and node repair

For the case where the linear codes $\mathcal{C}_1$ and $\mathcal{C}_2$ are MDS codes over $\mathbb{F}_q$, the data collector can perform the reconstruction process of the entire message only with contacting any $k$ nodes. All connected nodes must be of same type. The amount of stored data that will be downloaded during this process is $k^2 = S_{\mathrm{TW}}$. In general, it is important to note that the connectivity in such Twin-framework system must be at least $2k-1$ for satisfactory availability and higher guarantees.

In the case of a failed node, the newcomer from certain type must contact any $k$ nodes belonging to the opposite type. To recover the lost data just a single symbol from each node will be downloaded, that is $\beta = 1$. For a successful repair

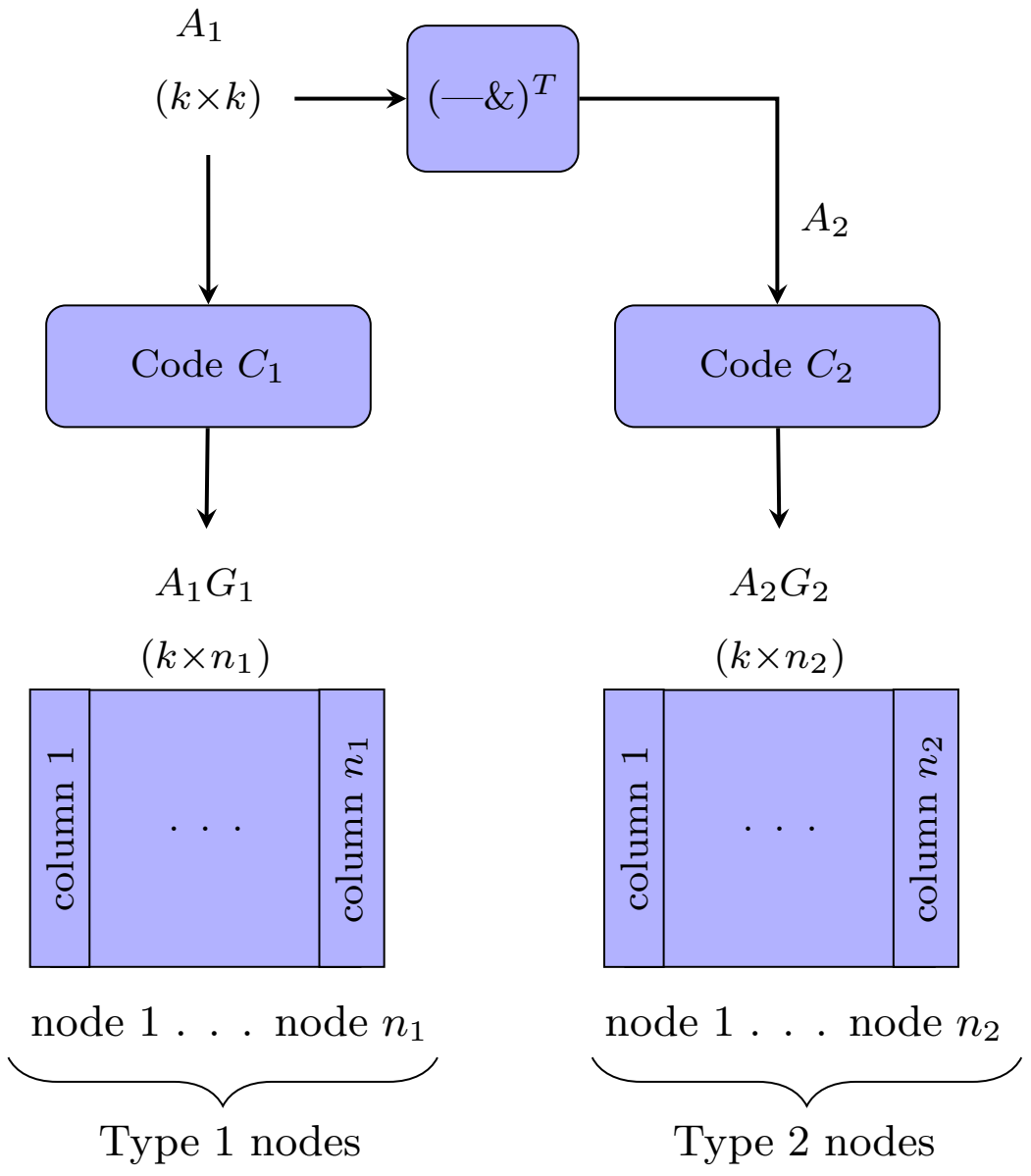

Figure 2: Encoding within the Twin-code framework.

process, in any moment, $k$ nodes from Type 1 and $k$ nodes from the Type 2 must be alive.

For example, if we assume that node $m$ from Type 1 fails, the newcomer (replacement node) must recover the following $k$ symbols $A_1\mathbf{g}_{(1,m)}$. Therefore, the newcomer contacts $k$ helper nodes of Type 2. The $j_r$–th helper node ($1 \leq j_r \leq n_2$) for all $r$, $1 \leq r \leq k$, sends the product of the encoding vector $\mathbf{g}_{(1,m)}$with the $k$ symbols of the helper node $A_2\mathbf{g}_{(2,j_r)}$, i.e. $\mathbf{g}_{(1,m)}^T A_2\mathbf{g}_{(2,j_r)}$. So, the replacement node obtains access to the $k$ symbols

$$\mathbf{g}_{(1,m)}^T A_2 \left[\mathbf{g}_{(2,j_1)} \dots \mathbf{g}_{(2,j_k)}\right]. \tag{6}$$

Defining

$$\boldsymbol{\tau}^T \triangleq \mathbf{g}_{(1,m)}^T A_2, \tag{7}$$

the newcomer has access to

$$\boldsymbol{\tau}^T \left[\mathbf{g}_{(2,j_1)} \dots \mathbf{g}_{(2,j_k)}\right] \tag{8}$$

and $\boldsymbol{\tau}^T$ is recovered by erasure decoding of the MDS code $\mathcal{C}_2$.

Therefore, the $k$ symbols that have to be recovered at the newcomer are the symbols contained into the vector

$$A_1\mathbf{g}_{(1,m)} = \left(\mathbf{g}_{(1,m)}^T A_2\right)^T = \boldsymbol{\tau}. \tag{9}$$

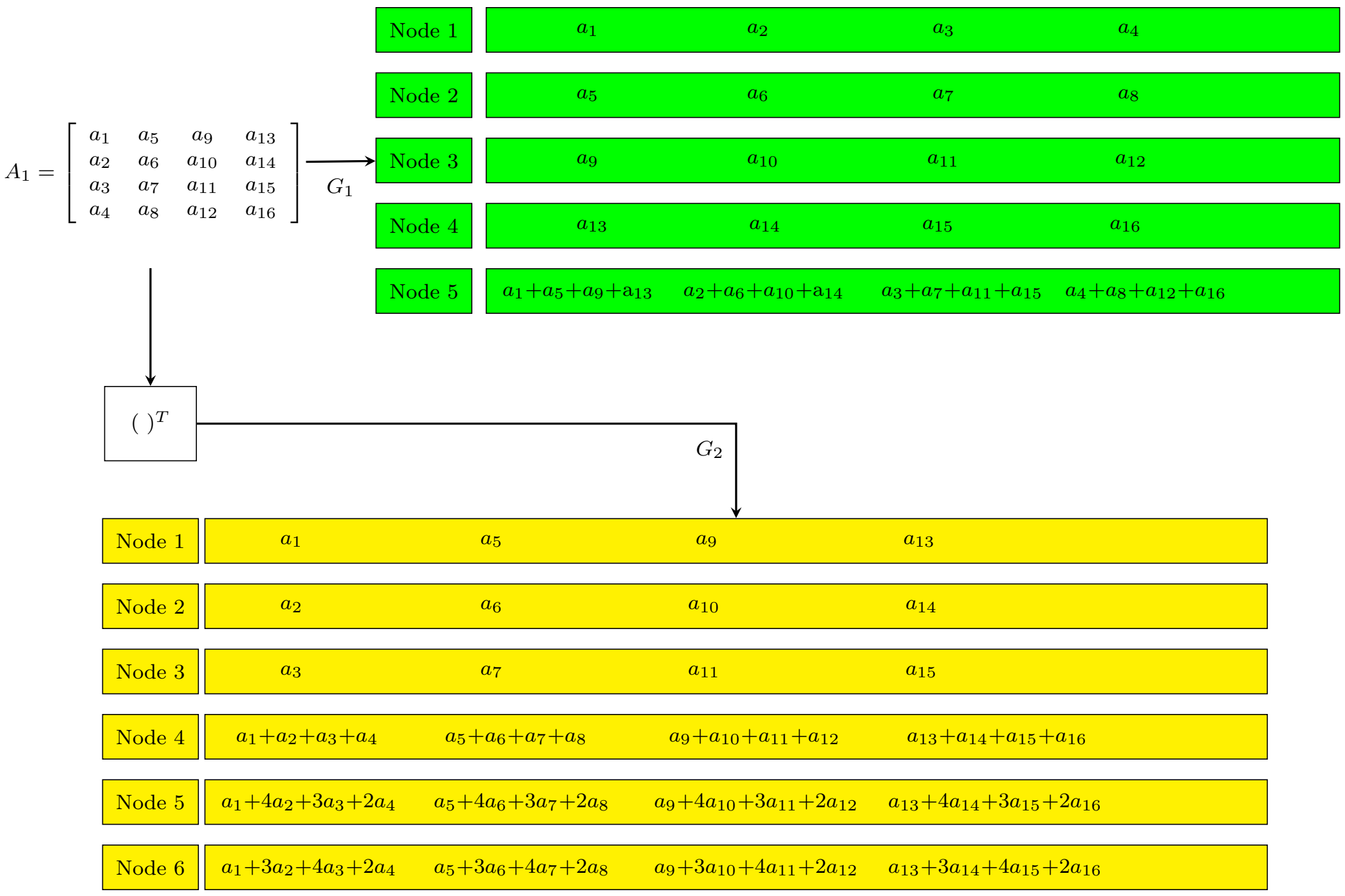

Figure 3: An example of distributing information within the Twin-code framework with parameters $n_1 = 5$, $n_2 = 6$ and $k = 4$.

Thus, the repair process of a Type 1 node is brought to the erasure decoding of the code $\mathcal{C}_2$.

*Example 1*: We illustrate the encoding in Twin-code framework with the following example. It will be used to discuss the security in the Twin-code framework. Let distribute the message of $S_{\text{TW}} = 16$ symbols arranged in message matrix

$$A_1 = \begin{bmatrix} a_1 & a_5 & a_9 & a_{13} \\ a_2 & a_6 & a_{10} & a_{14} \\ a_3 & a_7 & a_{11} & a_{15} \\ a_4 & a_8 & a_{12} & a_{16} \end{bmatrix}$$

in the network with $n_1 = 5$ nodes of Type 1 and $n_2 = 6$ nodes of Type 2 shown in Fig. 3. The Twin-MDS code generator matrices over $\mathbb{F}_{11}$ are

$$G_1 = \begin{bmatrix} 1 & 0 & 0 & 0 & 1 \\ 0 & 1 & 0 & 0 & 1 \\ 0 & 0 & 1 & 0 & 1 \\ 0 & 0 & 0 & 1 & 1 \end{bmatrix}, G_2 = \begin{bmatrix} 1 & 0 & 0 & 1 & 1 & 1 \\ 0 & 1 & 0 & 1 & 4 & 3 \\ 0 & 0 & 1 & 1 & 3 & 4 \\ 0 & 0 & 0 & 1 & 2 & 2 \end{bmatrix}.$$

The data collector can reconstruct the information by contacting any $k = 4$ nodes of the same type.

## 2.3 Advantages of the Twin – code framework in data reconstruction and repair processes

Using the Twin method, all encoding operations, data reconstruction and repair processes employ encoding and decoding algorithms according to the appropriate code. This makes the Twin framework method robust. Employing the existing code in the Twin-code framework is one of the main advantages. This feature allows utilization of any linear erasure code. Moreover, regarding the repair process only one of the erasure codes will be used, which reduces the complexity of the decoding process. Another positive impact is the reduced repair overhead. The repair algorithm in a Twin-code network is such that the symbols passed through the helper node are identity independent from the other nodes that help in the repair process. In other words, the replacement node encoding vector does not depend on the encoding vectors of the helper nodes.

Furthermore, Twin–code framework is very efficient in a data deployment in distributed storage network, since the data reconstruction and repair process are simple to be accomplished. This process starts with the source transmitting the encoded data to a subset of nodes, which means some of the nodes in the system stay empty. Then the empty nodes can be treated as replacement nodes and the process of data deployment will be finished with the help of their adjacent coded nodes as depicted in Fig. 4. This feature makes the traffic more uniform across the network. Finally, concerning the error detection and correction, the Twin-code network will use only the appropriate error correcting code from the constituent codes to resolve the error detection and correction problems.

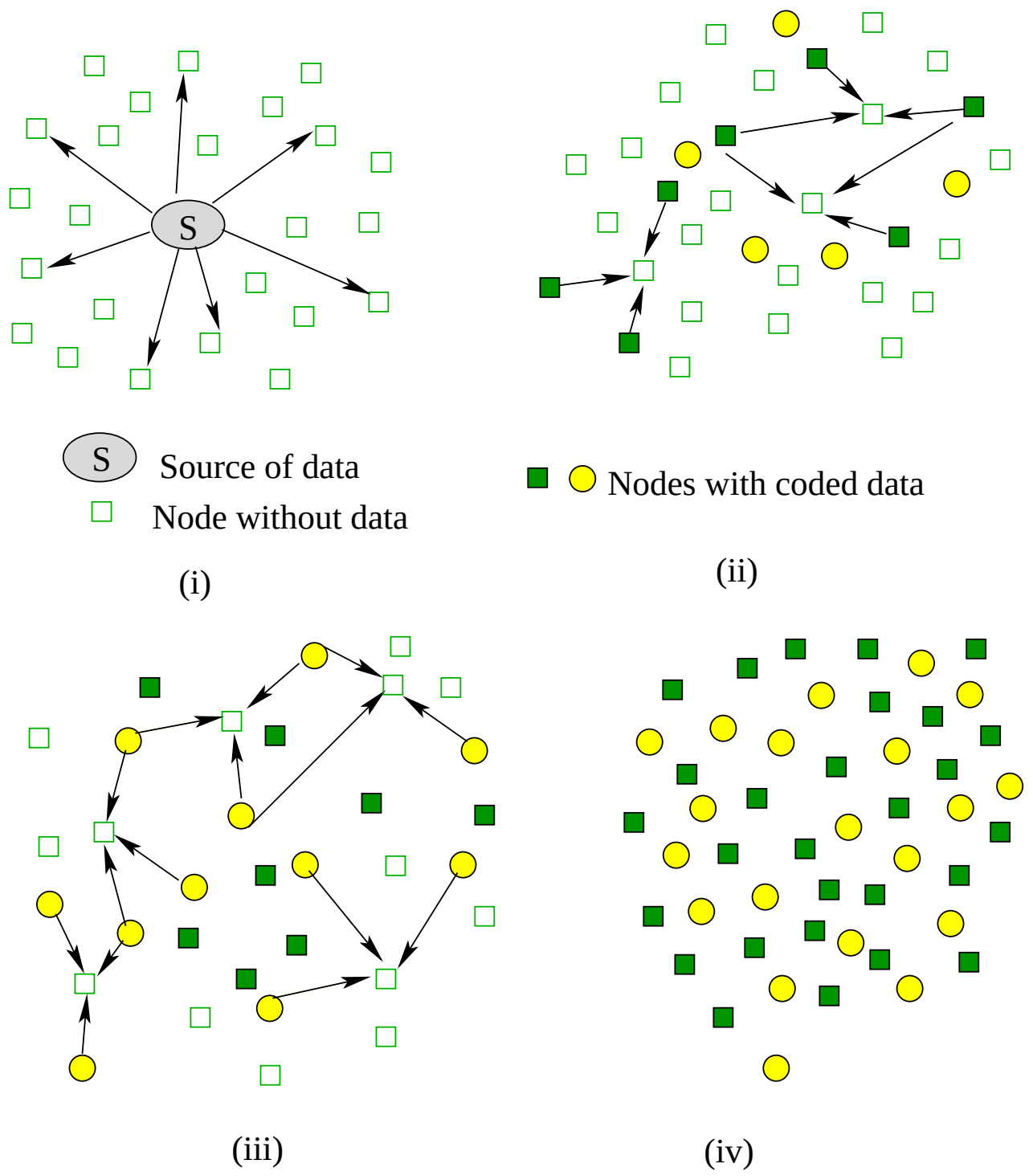

Figure 4: Data deployment in Twin-code distributed storage network: (i) Source of data distributes coded data to a subset of nodes. (ii), (iii) These live nodes help in the repair process of nodes that have lost their data. (iv) Final state of the system.

# 3 Comparison of data–reconstruction and node repair in DSS within regenerating codes and in Twin-code framework

In this section we compare the upper bounds of the number of message symbols that can be distributed by MBR and MSR codes with the maximum number of message symbols distributed within Twin MDS code framework and we show that the Twin MDS code framework gives better performance in a distributed file size than MBR codes and same performance as MSR codes.

The repair process within regenerating codes is established by downloading $\beta$ symbols from any subset of the remaining $d$ $(n \geq d \geq k)$ nodes. The total repair bandwidth $d\beta$ is usually smaller than the size of the message $S$. In [7], authors also establish that the parameters satisfy the bound

$$S \leq \sum_{i=0}^{k-1} \min\left(\alpha, \left(d-i\right)\beta\right). \tag{10}$$

MBR codes achieve minimal possible repair bandwidth $d\beta = \alpha$, i.e., the node downloads only what it stores. Plugging $d\beta = \alpha$ in (10), and replacing the inequality with equality, an MBR code, with no secrecy requirements, must satisfy

$$S = \left(kd - \left(\begin{array}{c} k \\ 2 \end{array}\right)\right)\beta. \tag{11}$$

Since in the Twin MDS code framework the repair bandwidth is $\beta = 1$, we compare it with MBR codes with the same bandwidth. MBR codes with $\beta = 1$ have data storage $\alpha = d$ and are constructed such that the message matrix is populated by $S$ message symbols

$$S = \left(kd - \left(\begin{array}{c} k \\ 2 \end{array}\right)\right) = k\left(d-k\right) + \frac{k\left(k+1\right)}{2}. \tag{12}$$

MSR codes achieve minimum possible storage at each node. Knowing that the message size is $S$, each node stores $\alpha = \frac{S}{k}$ data. For message reconstruction, the data-collector contacts any $k$ nodes. Based on that statement, from (10) and replacing the inequality with equality when there is no secrecy requirement, MSR codes must satisfy

$$S = k\alpha \quad \text{and} \quad d\beta = \alpha + \left(k-1\right)\beta. \tag{13}$$

Observing the case for code construction when $\beta = 1$ the equation (13) becomes

$$d = \alpha + \left(k-1\right). \tag{14}$$

Based on above stated further is made comparison between the data distribution process performed by the regenerating codes and the Twin MDS code framework. In the Twin MDS code framework, by definition, the size of the message that can be distributed in the system is maximum $S_{\mathrm{TW}} = k^2$. When the size of the message is larger than $k^2$ symbols, the message first is divided into fragments of size $k^2$. In that case, the method is applied to each of these fragments and at the end all of them are concatenated. Plugging $\beta = 1$, $\alpha = d = k$ in (12), the size of the distributed file with MBR codes becomes

$$S = \frac{k\left(k+1\right)}{2}. \tag{15}$$

Comparing the message size from (15) and the size of the message in the Twin MDS code framework $S_{\mathrm{TW}}$, we notice that the number of message symbols that can be distributed in the Twin MDS code storage system is greater or equal to the number of message symbols distributed by MBR codes in the special case when $\beta = 1$ and $\alpha = d = k$, i.e.,

$$S_{\text{TW}} = k^2 \geq S = \frac{k\,(k+1)}{2}.$$

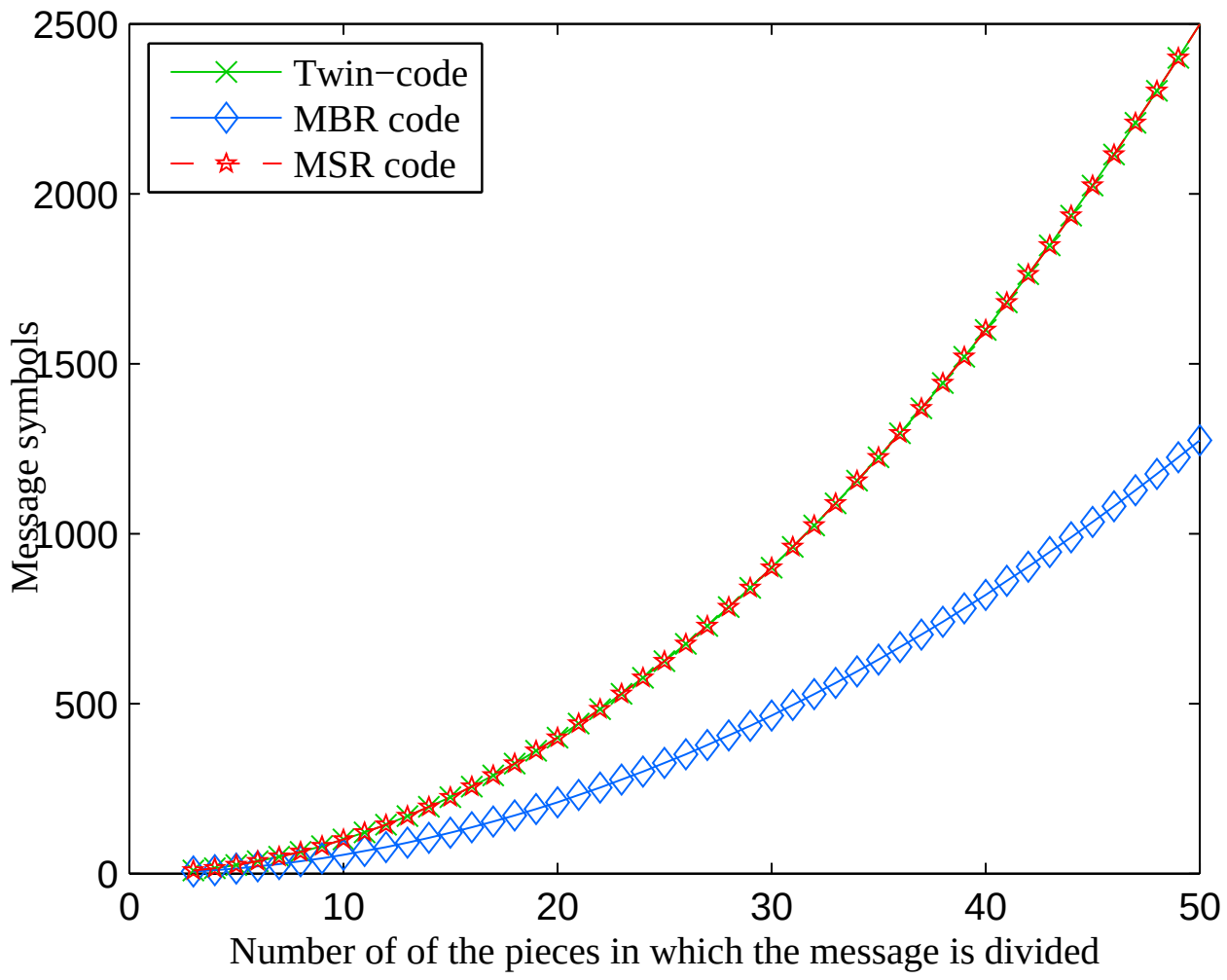

Figure 5: Comparison of the size of the message that can be distributed in Twin MDS code framework (times signs) and in DSS with MBR code (diamonds) and MSR code (stars) for $k = 3, ..., 50$ pieces in which the message is divided.

Since in the Twin MDS codes the connectivity must be at least $d = 2k - 1$, for the comparison of the file size distributed in Twin MDS code and the file size in the DSS with MSR codes we plugging $d = 2k - 1$ in (14). Therefore, the node storage should be $\alpha = k$ and the size of the distributed file with MSR codes is $S = k^2$. This indicates that the number of message symbols that can be distributed in the Twin MDS code storage system is the same as the number of message symbols distributed by MSR codes when $\beta = 1$ and $d = 2k - 1$, i.e., $S_{\text{TW}} = k^2 = S = k^2$.

In Fig. 5. we compare the size of the message symbols stored in the Twin MDS code framework and DSS with MBR and MSR codes. The size of the message symbols in all storage systems increases as the number of the number of the pieces $k$ in which the message is divided increases. However, the file size when $\beta = 1$ and $\alpha = d = k$ that supports the Twin MDS code framework is greater than the file size supported by MBR codes, and when $\beta = 1$, $d = 2k - 1$ the file size distributed in Twin MDS code framework is equal with the file size supported by MSR codes.

# 4 Secrecy in DSS

In this section we consider the upper bounds of the achievable secure file size that can be distributed in a network by MBR and MSR codes in presence of passive adversary. Next in the following subsections respectively, we construct a new secure Twin MDS code framework for which we calculate the number of message symbols that can be securely stored in. In addition, we prove that the achievable secure file size stored in a network by MBR and MSR codes is less than the stored secure file size in this new Twin MDS code framework.

We consider an $(l_1, l_2)$ eavesdropper model, $l = l_1 + l_2 < k$, defined in [15] where an eavesdropper may gain access to: either the data stored in a subset of $l_1$ storage nodes, where the set of eavesdropped indices is denoted by $\mathcal{E}_1$, or to the data downloaded during the repair process of other $l_2$ nodes, where the set of these observed indices is denoted by $\mathcal{E}_2$, or both. This concept is a generalized version from the eavesdropper model considered by Pawar et al. in [13] and [14]. In [13] an eavesdropper may gain access only to the data stored on the nodes, which indicates to the MBR point. When the repair bandwidth is strictly greater than the per node storage, then is considered the MSR point, meaning an eavesdropper gains more information if it has access to the data downloaded during node repair if compared to the case when it observes only the data stored on the node. The achievability of security in this $(l_1, l_2)$ eavesdropper model is formalized by the following definition:

**Definition 1 (Security in an $(l_1, l_2)$ Eavesdropper [15])** *Consider a DSS in which an eavesdropper gains access to the data stored on some $l_1$ nodes, and the data downloaded during repair on some other $l_2$ nodes. An $(l_1, l_2)$ secure distributed storage system with secure file size $S^{(s)}$ is such, where an eavesdropper obtains no information about the message, i.e. $I(f^s; \mathbf{e}) = 0$, where $f^s$ is the secure information of size $S^{(s)}$, and $\mathbf{e}$ represents the eavesdropper's observation.*

In [13], Pawar et al. provided an upper bound to the number of message symbols $S^{(s)}$ that can securely be stored in the system in presence of $l$ eavesdroppers. The bound is given by

$$S^{(s)} \leq \sum_{i=l}^{k-1} \min\left(\alpha, (d-i)\,\beta\right). \tag{16}$$

Since here we discuss about the exact repair for the MBR codes $d\beta = \alpha$, the replacement node downloads only the original stored data. In this case the eavesdropper cannot obtain any extra downloaded information from the repair process. Thus, without loss of generality it may be assumed that $l_2 = 0$. Therefore, the upper bound for secure MBR codes from the equation (16) can be obtained with substituting $\alpha = d\beta$ and replacing the inequality with equality. The upper bound of the MBR codes is

$$S^{(s)} = \left(kd - \binom{k}{2}\right)\beta - \left(ld - \binom{l}{2}\right)\beta. \tag{17}$$

Type 1 nodes

| Node | Node | Node | Node | Node 5 |
|---|---|---|---|---|
| $r_1$ | $r_5$ | $a_9$ | $a_{13}$ | $r_1 + r_5 + a_9 + a_{13}$ |
| $r_2$ | $r_6$ | $a_{10}$ | $a_{14}$ | $r_2 + r_6 + a_{10} + a_{14}$ |
| $r_3$ | $r_7$ | $a_{11}$ | $a_{15}$ | $r_3 + r_7 + a_{11} + a_{15}$ |
| $r_4$ | $r_8$ | $a_{12}$ | $a_{16}$ | $r_4 + r_8 + a_{12} + a_{16}$ |

Type 2 nodes

| Node | Node | Node | Node 4 | Node 5 | Node 6 |
|---|---|---|---|---|---|
| $r_1$ | $r_2$ | $r_3$ | $r_1 + r_2 + r_3 + r$ | $r_1 + 4r_2 + 3r_3 +$ | $r_1 + 3r_2 + 4r_3 + 2r_4$ |
| $r_5$ | $r_6$ | $r_7$ | $r_5 + r_6 + r_7 + r$ | $r_5 + 4r_6 + 3r_7 +$ | $r_5 + 3r_6 + 4r_7 + 2r_8$ |
| $a_9$ | $a_{10}$ | $a_{11}$ | $a_9 + a_{10} + a_{11} +$ | $a_9 + 4a_{10} + 3a_{11} +$ | $a_9 + 3a_{10} + 4a_{11} + 2a_{12}$ |
| $a_{13}$ | $a_{14}$ | $a_{15}$ | $a_{13} + a_{14} + a_{15} +$ | $a_{13} + 4a_{14} + 3a_{15} +$ | $a_{13} + 3a_{14} + 4a_{15} + 2a_{16}$ |

(i)

Type 1 nodes

| Node | Node | Node | Node | Node 5 |
|---|---|---|---|---|
| $r_1$ | $r_5$ | $a_9$ | $a_{13}$ | $r_1 + r_5 + a_9 + a_{13}$ |
| $r_2$ | $r_6$ | $a_{10}$ | $a_{14}$ | $r_2 + r_6 + a_{10} + a_{14}$ |
| $r_3$ | $r_7$ | $a_{11}$ | $a_{15}$ | $r_3 + r_7 + a_{11} + a_{15}$ |
| $r_4$ | $r_8$ | $a_{12}$ | $a_{16}$ | $r_4 + r_8 + a_{12} + a_{16}$ |

Type 2 nodes

| Node | Node | Node | Node 4 | Node 5 | Node 6 |
|---|---|---|---|---|---|
| $r_1$ | $r_2$ | $r_3$ | $r_1 + r_2 + r_3 + r_4$ | $r_1 + 4r_2 + 3r_32 + 2r$ | $r_1 + 3r_2 + 4r_32 + 2r_4$ |
| $r_5$ | $r_6$ | $r_7$ | $r_5 + r_6 + r_7 + r_8$ | $r_5 + 4r_6 + 3r_7 + 2r$ | $r_5 + 3r_6 + 4r_7 + 2r_8$ |
| $a_9$ | $a_{10}$ | $a_{11}$ | $a_9 + a_{10} + a_{11} + a_1$ | $a_9 + 4a_{10} + 3a_{11} + 2a$ | $a_9 + 3a_{10} + 4a_{11} + 2a_{12}$ |
| $a_{13}$ | $a_{14}$ | $a_{15}$ | $a_{13} + a_{14} + a_{15} + a_1$ | $a_{13} + 4a_{14} + 3a_{15} + 2$ | $a_{13} + 3a_{14} + 4a_{15} + 2a_{16}$ |

(ii)

Figure 6: Security in Twin $(5, 4)$ MDS - $(6, 4)$ MDS code in presence of two eavesdroppers: (i) Two nodes from different types are eavesdropped; (ii) Two nodes from same type, Type 1 are eavesdropped.

For MSR codes, the eavesdropper has an access to $l_1$ nodes, and listens $l_2$ nodes that are in the reparation process. In [20] Goparaju et al. have established an upper bound of the achievable secure file size in presence of both types of attack,

$$S^{(s)} = (k - l_1 - l_2)\left(1 - \frac{1}{d - k + 1}\right)^{l_2} \alpha. \tag{18}$$

Therefore, to obtain secure coding schemes at the MBR and MSR point that have better rate and/or secrecy capacity than that of the schemes proposed in [15] and [20] is a challenge.

In this section, we define a new secure Twin MDS framework in presence of an $(l_1, l_2)$ eavesdropper and we show that this framework gives better secrecy performance than MBR codes when $\beta = 1$, $\alpha = d = k$ and $l_2 = 0$, $(l_1 < k)$ and than the MSR codes when $\beta = 1$, $d = 2k - 1$ and $l = l_1 + l_2$ $(l < k)$. In the paper we use the following lemma to show that the proposed framework satisfy the secrecy constraints.

**Lemma 1 (Secrecy Lemma [15])** *Consider a system with secure information* $\mathbf{f^s}$*, random symbols* $\mathbf{r}$ *(independent of* $\mathbf{f^s}$*), and an eavesdropper with observations given by* $\mathbf{e}$*. If* $H(\mathbf{e}) \leq H(\mathbf{r})$ *and* $H(\mathbf{r}|\mathbf{f^s}, \mathbf{e}) = 0$*, then the mutual information leakage to eavesdropper is zero, i.e.,* $I(\mathbf{f^s}; \mathbf{e}) = 0$.

$$
\begin{aligned}
I(\mathbf{f^s};\mathbf{e}) &= H(\mathbf{e}) - H(\mathbf{e}|\mathbf{f^s}) \\
&\overset{(a)}{\leq} H(\mathbf{e}) - H(\mathbf{e}|\mathbf{f^s}) + H(\mathbf{e}|\mathbf{f^s},\mathbf{r}) \\
&\overset{(b)}{\leq} H(\mathbf{r}) - I(\mathbf{e};\mathbf{r}|\mathbf{f^s}) \\
&\overset{(c)}{=} H(\mathbf{r}|\mathbf{f^s},\mathbf{e}) \\
&\overset{(d)}{=} 0
\end{aligned}
$$

$(a)$ follows by non-negativity of $H(\mathbf{e}|\mathbf{f^s},\mathbf{r})$, $(b)$ is the condition $H(\mathbf{e}) \leq H(\mathbf{r})$, $(c)$ is due to $H(\mathbf{r}|\mathbf{f^s}) = H(\mathbf{r})$ as $\mathbf{r}$ and $\mathbf{f^s}$ are independent, $(d)$ is the condition $H(\mathbf{r}|\mathbf{f^s},\mathbf{e}) = 0$.

## 4.1 Secure Twin MDS code framework

In this subsection, we present explicit construction for a coding scheme that is secure against an $(l_1, l_2)$ eavesdropper when $\mathcal{E}_1 \cup \mathcal{E}_2 \subset \mathcal{E}$, for a given set $\mathcal{E}$ of size $|\mathcal{E}| < k$, all parameter values $[n, k, d = k]$ and $\alpha = k$, $\beta = 1$. The constructions is based on Twin MDS codes such that the message matrix first is modified with $S_{\text{TW}} - S_{\text{TW}}^{(s)}$ random symbols, where $S_{\text{TW}}^{(s)}$ is the number of message symbols that can be securely stored in a Twin MDS framework in an $(l_1, l_2)$ eavesdropper model.

First, we state the following property associated with the repair process in a Twin-code framework.

**Lemma 2**

1. *Assume that an eavesdropper gains access to the data stored on $l = l_1$ nodes in a Twin MDS code framework. Then the eavesdropper can only observe $lk$ independent symbols.*

2. *Assume that an eavesdropper has an access to the data stored on any $l_1$ nodes and observes the downloaded data from $l_2$ nodes that are in reparation process in a Twin MDS code framework. Then the $(l_1, l_2)$ eavesdropper can observe at most $k(l_1 + l_2)$ independent symbols.*

1. Since the size of the stored data on each node is $k$ symbols by the construction of Twin MDS code the maximum number of independent symbols that the intruder can reveal is $lk$ if $\mathcal{E}_1 \subset [k]^1$ or $\mathcal{E}_2 \subset [k]^2$ , i.e., when it gains access in the data stored on $l_1$ nodes of the same type.

2. From Lemma 2, 1) in an $(l_1, l_2)$ eavesdropper model, $(l_1 + l_2) < k$ the intruder can only observe $kl_1$ independent symbols when it gains access in the data stored on $l_1$ nodes. Since in Twin MDS code framework the repair bandwidth is $k$, i.e., a newcomer node of a certain type can recover the symbols stored in the failed node by downloading a single symbol from any $k$ nodes of the other type, the maximum number of independent symbols that the intruder can reveal is $kl_2$ if $\mathcal{E}_2 \subset N_1$ or $\mathcal{E}_2 \subset N_2$ when it gains access the downloaded data of $l_2$ failed systematic nodes of same type. Therefore, the maximum number of message symbols that the intruder can reveal if it can read-access the data stored in $l_1$ nodes and read-access the downloaded data during the repair process of $l_2$ failed systematic nodes is $k(l_1 + l_2)$.

Now, we detail an achievability scheme of this section. There are $n_1$ Type 1 nodes, and $n_2$ Type 2 nodes, where $n = n_1 + n_2$ is the total number of storage nodes in the framework. Note that the storage nodes of both types have the same characteristics. In the sequel we use the following notations:

- $[k]$ is a set of indices of any $k$ repair nodes (nodes involved in repair process).
- $[k]^1$ is a set of indices of any $k$ nodes of Type 1 and $[k]^2$ is a set of indices of any $k$ nodes of Type 2.
- $N_1$ ($|N_1| \leq k$) is a set of indices of systematic nodes of Type 1 and $N_2$ ($|N_2| \leq k$) is a set of indices of systematic nodes of Type 2.

Let $\mathbf{f^s}$ is secure information of size $S_{\text{TW}}^{(s)} = k(k-l)$ or $S_{\text{TW}}^{(s)} = k(k-l_1-l_2)$ in $\mathbb{F}_q$ at MBR and MSR points, respectively, i.e., $\mathbf{f^s} = (a_1, a_2, ..., a_{k(k-l)})$ or $\mathbf{f^s} = (a_1, a_2, ..., a_{k(k-l_1-l_2)})$. We take $kl$ or $k(l_1+l_2)$ i.i.d. random symbols $\mathbf{r} = (r_1, ..., r_{kl})$ or $\mathbf{r} = (r_1, ..., r_{k(l_1+l_2)})$ at MBR and MSR points, respectively; distributed uniformly at random over $\mathbb{F}_q$, and append $\mathbf{r}$ to obtain $\mathbf{f} = (\mathbf{r}, \mathbf{f^s}) \in \mathbb{F}_q$, that will be encoded in the following manner:

- Arrange the message $\mathbf{f} = (f_1, ..., f_{k^2})$ into $(k \times k)$ matrix

$$A_1 = \begin{bmatrix} r_1 & \dots & r_{(l-1)k+1} & a_1 & a_{k+1} & \dots & a_{(k-l-1)k+1} \\ r_2 & \dots & r_{(l-1)k+2} & a_2 & a_{k+2} & \dots & a_{(k-l-1)k+2} \\ . & \dots & . & . & . & \dots & . \\ \\ r_k & \dots & r_{lk} & a_k & a_{2k} & \dots & a_{k(k-l)} \end{bmatrix}.$$

- Use the encoding algorithm defined in Section II. A.

With this algorithm the data is encoded and mapped into the network. Next, we present the following results of security for the general coding scheme described above.

**Theorem 1**

1. *The code based on Twin MDS code that is modifying the message matrix with $kl$ random symbols, explained as above achieves a secure file size $k(k-l)$ in an $(l_1, l_2)$ eavesdropper model, where $l_2 = 0$ and $l = l_1 < k$ at MBR point with $= k$ and $\beta = 1$.*
2. *The code based on Twin MDS code that is modifying the message matrix with $k(l_1 + l_2)$ random symbols, explained as above achieves a secure file size $k(k - l_1 - l_2)$ in an $(l_1, l_2)$ eavesdropper model, where $\mathcal{E}_2 \subset N_1 \cup N_2$, $l_1 + l_2 < k$ at MBR point with $= k$, $\beta = 1$ and $d = 2k - 1$.*

Type 1 nodes

| Node 1 | Node 2 | Node 3 | Node 4 | Node 5 |
|---|---|---|---|---|
| $r_1$ | $r_5$ | $a_9$ | $a_{13}$ | $r_1 + r_5 + a_9 + a_{13}$ |
| $r_2$ | $r_6$ | $a_{10}$ | $a_{14}$ | $r_2 + r_6 + a_{10} + a_{14}$ |
| $r_3$ | $r_7$ | $a_{11}$ | $a_{15}$ | $r_3 + r_7 + a_{11} + a_{15}$ |
| $r_4$ | $r_8$ | $a_{12}$ | $a_{16}$ | $r_4 + r_8 + a_{12} + a_{16}$ |

Type 2 nodes

| Node 1 | Node 2 | Node 3 | Node 4 | Node 5 | Node 6 |
|---|---|---|---|---|---|
| $r_1$ | | $r_3$ | $r_1 + r_2 + r_3 + r$ | $r_1 + 4r_2 + 3r_3 + 2$ | $r_1 + 3r_2 + 4r_3 + 2r_4$ |
| $r_5$ | | $r_7$ | $r_5 + r_6 + r_7 + r$ | $r_5 + 4r_6 + 3r_7 + 2$ | $r_5 + 3r_6 + 4r_7 + 2r_8$ |
| $a_9$ | | $a_{11}$ | $a_9 + a_{10} + a_{11} + a$ | $a_9 + 4a_{10} + 3a_{11} +$ | $a_9 + 3a_{10} + 4a_{11} + 2a_{12}$ |
| $a_{13}$ | | $a_{15}$ | $a_{13} + a_{14} + a_{15} +$ | $a_{13} + 4a_{14} + 3a_{15} +$ | $a_{13} + 3a_{14} + 4a_{15} + 2a_{16}$ |

(i)

Type 1 nodes

| Node 1 | Node 2 | Node 3 | Node 4 | Node 5 |
|---|---|---|---|---|
| $r_1$ | $r_5$ | $a_9$ | $a_{13}$ | $r_1 + r_5 + a_9 + a_{13}$ |
| $r_2$ | $r_6$ | $a_{10}$ | $a_{14}$ | $r_2 + r_6 + a_{10} + a_{14}$ |
| $r_3$ | $r_7$ | $a_{11}$ | $a_{15}$ | $r_3 + r_7 + a_{11} + a_{15}$ |
| $r_4$ | $r_8$ | $a_{12}$ | $a_{16}$ | $r_4 + r_8 + a_{12} + a_{16}$ |

Type 2 nodes

| Node 1 | Node 2 | Node 3 | Node 4 | Node 5 | Node 6 |
|---|---|---|---|---|---|
| $r_1$ | $r_2$ | $r_3$ | $r_1 + r_2 + r_3 + r_4$ | $r_1 + 4r_2 + 3r_32 + 2r$ | $r_1 + 3r_2 + 4r_32 + 2r_4$ |
| $r_5$ | $r_6$ | $r_7$ | $r_5 + r_6 + r_7 + r_8$ | $r_5 + 4r_6 + 3r_7 + 2r$ | $r_5 + 3r_6 + 4r_7 + 2r_8$ |
| $a_9$ | $a_{10}$ | $a_{11}$ | $a_9 + a_{10} + a_{11} + a_1$ | $a_9 + 4a_{10} + 3a_{11} + 2a$ | $a_9 + 3a_{10} + 4a_{11} + 2a_{12}$ |
| $a_{13}$ | $a_{14}$ | $a_{15}$ | $a_{13} + a_{14} + a_{15} + a_1$ | $a_{13} + 4a_{14} + 3a_{15} + 2$ | $a_{13} + 3a_{14} + 4a_{15} + 2a_{16}$ |

(ii)

Figure 7: $(l_1, l_2)$=(1,1) eavesdropper at the first and second node of Type 2 nodes, respectively in Twin (5,4) MDS - (6,4) MDS code framework. (i) Encoded symbols obtained by an eavesdropper by observing repair of node 2 of Type 2. (ii) Encoded symbols obtained by eavesdropping the first node from Type 2 and observing repair of the second node from Type 2.

1. The repair and data reconstruction properties of the proposed code follow from the construction code in [12]. We use Lemma 1 to prove the security of this code against an $(l_1, l_2)$ eavesdropper with $l_2 = 0$, $l = l_1$. Considering that $\mathbf{e}$ denotes the symbols observed by an eavesdropper, we need to show: (i) $H(\mathbf{e}) \leq H(\mathbf{r})$ and (ii) $H(\mathbf{r}|\mathbf{f^s}, \mathbf{e}) = 0$. It follows from Lemma 2, 1) that an eavesdropper observes $kl$ independent symbols, and since $|\mathbf{e}| = kl$, follows that $H(\mathbf{e}) = H(\mathbf{r})$, which is the first requirement for establishing the security claim. It remains $H(\mathbf{r}|\mathbf{f^s}, \mathbf{e}) = 0$, i.e., to show that given the message symbols as side information, an eavesdropper can decode all the random symbols. To this end, without loss of generality we assume that the eavesdropper gain access to the data stored on $l$ nodes of Type 1. Now, define $A_1^{(s)}$ as a $(k \times k)$ matrix obtained by setting all symbols of the secure file $\mathbf{f^s}$ in $A_1$ to zero. Thus $A_1^{(s)}$ has its first $l$ columns identical to that of $A_1$, and zeros elsewhere. Let $\tilde{\mathbf{e}} = A_1^{(s)} \left[\mathbf{g}_{(1,1)} \ldots \mathbf{g}_{(1,l)}\right]$ are the $lk$ symbols that the eavesdropper has access to, given the secure message symbols as side information. The MDS property of code $\mathcal{C}_1$ guarantees linear independence of the corresponding $l$ columns of generator matrix $\left[\mathbf{g}_{(1,1)} \ldots \mathbf{g}_{(1,l)}\right]$. So, recovering the random symbols $\mathbf{r}$ from $\tilde{\mathbf{e}}$ is identical to data reconstruction in the original $\tilde{\mathcal{C}}_1$ code designed for $(n_1, k = l)$ and no eavesdroppers. Thus, given the secure message symbols, the eavesdropper can decode all the random symbols, i.e., $H(\mathbf{r}|\mathbf{f^s}, \mathbf{e}) = 0$.

2. Similar as the proof of Theorem 1, 1).

The security achievement of the Twin MDS code framework applied in Example 1 is presented in Fig. 6. In this scenario we consider the passive type $l_1$ attack. The number of nodes that are compromised is $l_1 = 2$. In Fig. 6 (i), one compromised node belongs to the nodes of Type 1 and the other one is of Type 2 nodes; more precisely Node 1 of Type 1 and Node 2 of Type 2 are affected. We note that the intruder reveals seven message symbols $\{r_1, r_2, r_3, r_4, r_6, a_{10}, a_{14}\}$.

That is less than the maximum number of message symbols that the intruder can reveal since it eavesdrops two nodes, i.e., $lk = 8$ compromised message symbols. Therefore, although the data collector can reconstruct the information, $l_1 = 2$ compromised nodes (less by assumption than $k = 4$) are not sufficient for the intruder to reveal the entire message. The situation for providing security in the system is similar for the situation in Fig. 6 (ii). The only difference here is that both compromised nodes belong to the same type of nodes: Node 2 and Node 3 of Type 1 nodes. In this case the intruder reveals the maximum number of message symbols when it eavesdrops two nodes, but still can not reveal the information which is distributed in the system.

*Example 2:* In this example, we illustrate the security in Twin MDS code framework in presence of eavesdroppers that can gain access both to stored data in $l_1 = 1$ node and to downloaded data during the repair of $l_2 = 1$ failed node. The information in the Twin MDS code framework is distributed same as in the system in Example 1. We assume that Node 2 of Type 2 nodes fails. The repair process of this failed node is shown in Fig. 7 (i), where the newcomer contacts helper nodes: Node 1, Node 3, Node 4 and Node 5 of Type 1 to recover the lost data. In Fig. 7 (ii) we present the revealed message symbols: $\{r_1, r_2, r_5, r_6, a_9, a_{10}, a_{12}, a_{14}\}$, if the eavesdropper can read the data stored on Node 1 of Type 2 and to the downloaded data during the repair process of the failed Node 2 of Type 2. Therefore, although the data collector can reconstruct the total information, the intruder can not reveal the entire message.

## 4.2 Information-theoretic secrecy in Twin MDS code framework and DSS with MBR codes

In this subsection we compare the secrecy in Twin MDS code framework and in storage system where the information is distributed with MBR codes in a presence of passive eavesdroppers. Since from the MBR point during the repair process the replacement node downloads only what it was stored we will consider only $l_1$ ($l_1 \leq k$) collaborating eavesdroppers which may gain access to the data stored but not the data downloaded during the repair process of some nodes, i.e., $l_2 = 0$.

**Corollary 2** *(from Theorem 1) The Twin MDS code framework gives better secrecy performance than MBR codes for $\beta = 1$ and $\alpha = d = k$ in presence of $l = l_1$ eavesdroppers, $(l < k)$.*

Plugging $\alpha = d = k$ in (17), the size of the secure message symbols achieved with MBR codes, when $\beta = 1$ is

$$S^{(s)} = \frac{k-l}{2} \cdot (k+1-l)\,. \qquad (19)$$

From Theorem 1 1) $S^{(s)}_{\mathrm{TW}} = k(k-l)$. So, since

$$k\,(k-l) \geq \frac{k-l}{2} \cdot (k+1-l)\ ,$$

follows that

$$S^{(s)}_{\mathrm{TW}} \geq S^{(s)}\ .$$

This means that the number of message symbols that can be securely stored in the Twin-code framework is greater or equal than the number of the secure message symbols stored in a system using MBR codes.

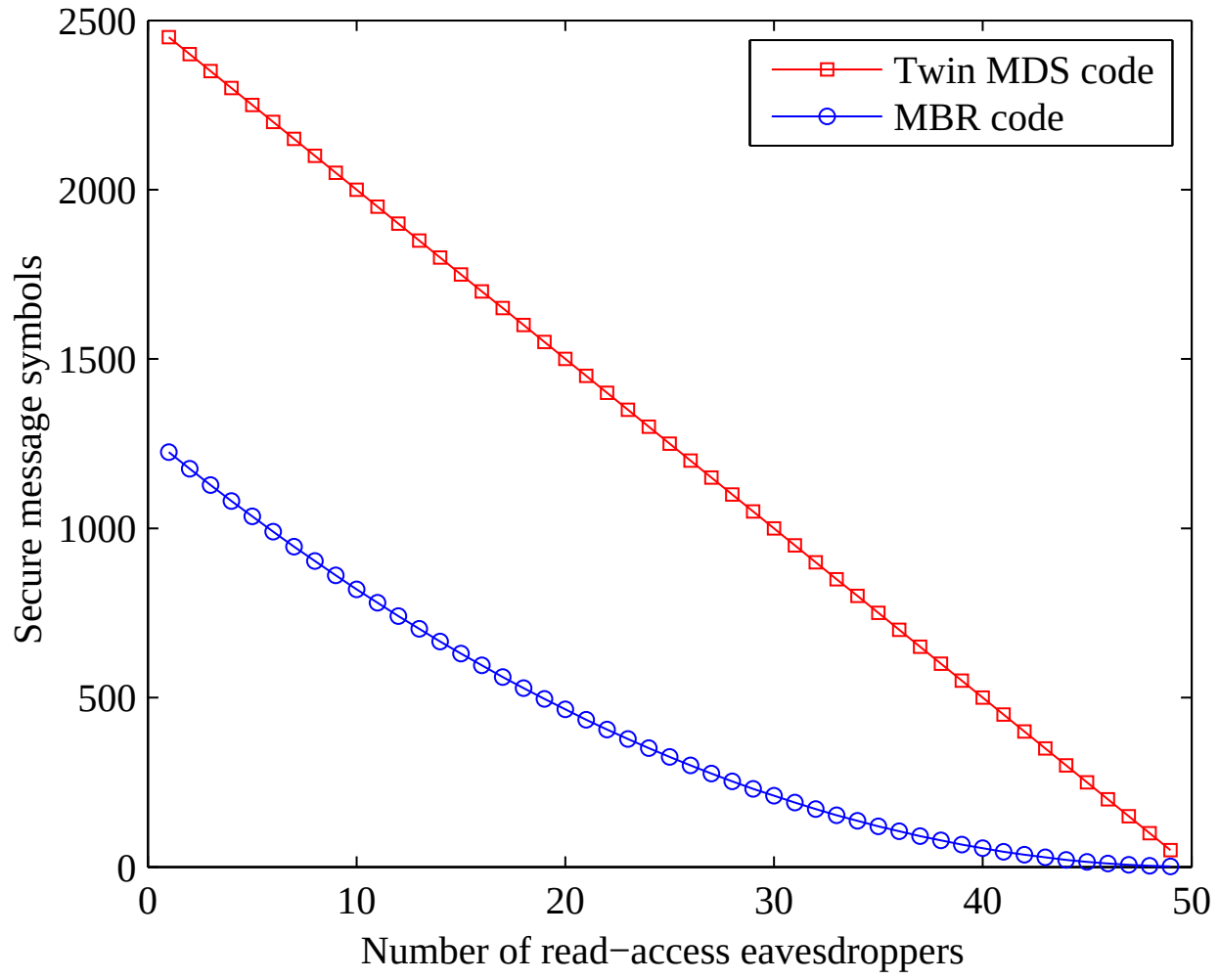

Figure 8: (a) Comparison of the size of the message that can be securely stored in Twin MDS code framework (squares) and MBR code (circles) for $k = 50$ in presence of $l = l_1 = 1, ..., 49$ eavesdroppers.

In Fig. 8 we compare the size of the secure message symbols in the Twin MDS code storage system (squares) and a DSS with MBR codes (circles). From this figure we notice that the size of the secure message symbols decreases in both Twin MDS and MBR codes distributed networks as the number of passive eavesdroppers increases. However, the secure file size in presence of $l$ eavesdroppers when $\beta = 1$ and $\alpha = d = k$ that supports the Twin MDS code framework is greater or equal to the secure file size supported by MBR codes. Fig. 8 is a graphical representation for the validity of Theorem 2.

## 4.3 Information-theoretic Secrecy in Twin MDS code framework and DSS with MSR codes

In [20], Goparaju et al. prove that the upper bound of the achievable secure file size $S^{(s)}$ for the given $(n, k, d)$, the MSR code, with $\alpha$ node storage capacity and $d$ contacted nodes during the repair process, in the situation when the eavesdropper has access to the data stored on $l_1$ nodes and to the downloaded data from $l_2$ systematic nodes is given by (18). Since in the Twin MDS codes the connectivity must be at least $d = 2k - 1$, for the comparison of the secure filze size distributed in Twin MDS code and the secure file size in the DSS with MSR codes we plugging $d = 2k - 1$ in (18), the secure size file becomes

$$S^{(s)} = \alpha \left(k - l_1 - l_2\right) \left(\frac{k-1}{k}\right)^{l_2}. \tag{20}$$

**Corollary 3** *(from Theorem 1) The Twin MDS code framework gives better secrecy performance than the MSR codes in a distributed storage system, when $\beta = 1$, $d = 2k - 1$ and $l = l_1 + l_2$ $(l < k)$ nodes are compromised.*

The node storage capacity in a distributed storage system that uses MSR codes with $d = 2k - 1$ is $\alpha = k$. Hence, plugging $\alpha = k$ in (20) the achievable secure file size in presence of both types of attacks is

$$S^{(s)} = k \left(k - l_1 - l_2\right) \left(\frac{k-1}{k}\right)^{l_2}.$$

From Theorem 1 2) in $(l_1, l_2)$ eavesdropper model $S^{(s)}_{\mathrm{TW}} = k(k - l_1 - l_2)$.

Since $1 > \frac{k-1}{k}$, then

$$1 > \left(\frac{k-1}{k}\right)^{l_2},$$

which implies that

$$S^{(s)}_{\mathrm{TW}} > S^{(s)}.$$

This means that the secure file stored in the Twin MDS code framework is larger in size than the secure file stored in a distributed storage system that uses MSR codes in presence of eavesdroppers of both types.

In Fig. 9 we compare the size of the secure message between the Twin MDS code storage system (dots) and the distributed storage system with MSR codes (triangles). Although the number of secure message symbols decreases in both cases, as the number of eavesdroppers that gain access to the data downloaded during repair of some failed nodes is increasing, the secure file size is larger for the Twin MDS code framework when $\beta = 1$ and $d = 2k-1$. Fig. 9 is a graphical representation for the validity of Theorem 3.

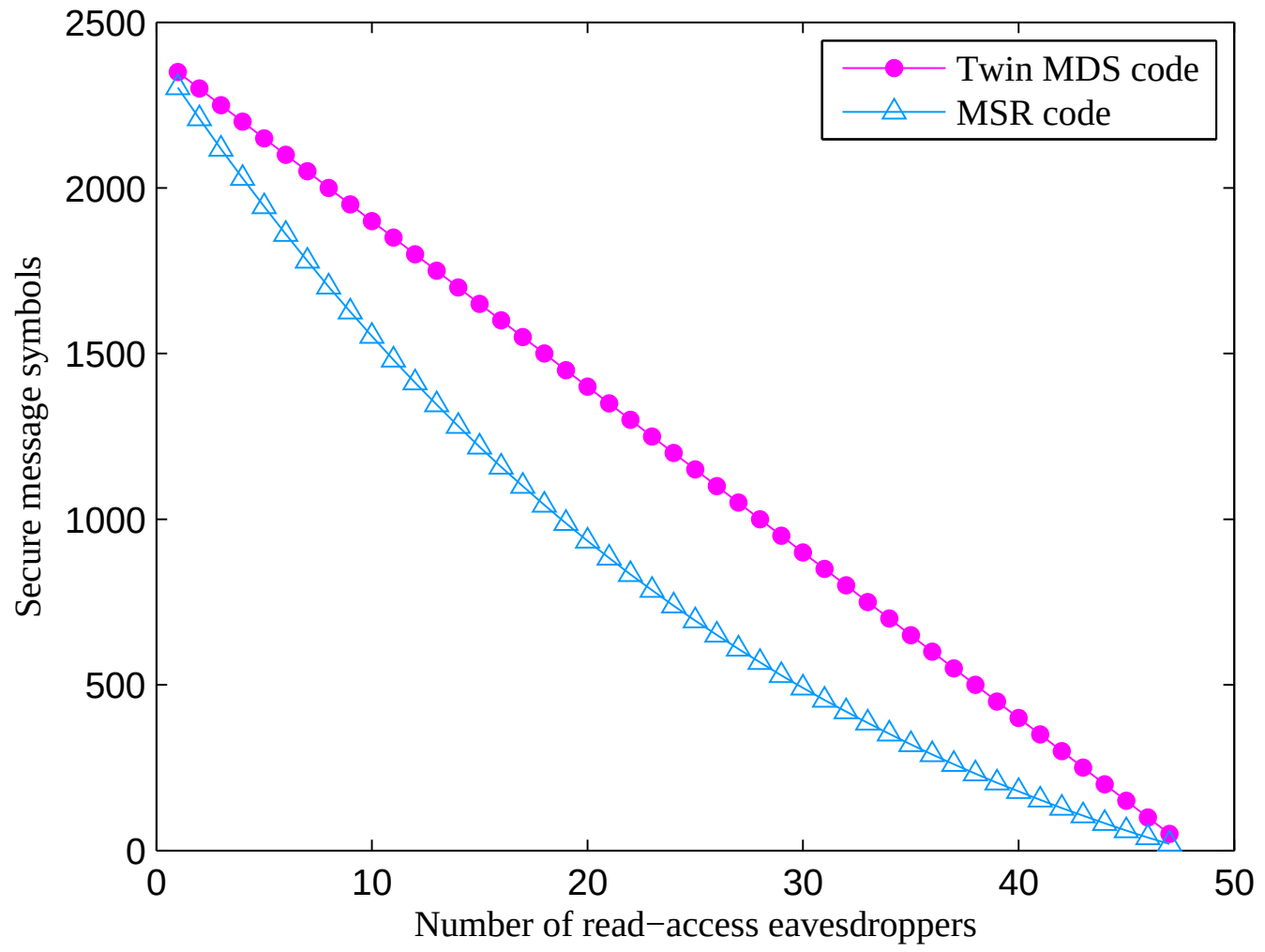

Figure 9: Comparison of the size of the message that can be securely stored in Twin MDS code framework (dots) and MSR code (triangles), for $k = 50$ in presence of $l_1 = 2$ and $l_2 = 1, ..., 47$ compromised nodes.

# 5 Conclusion

In this paper we considered reliable cloud storage system with more efficient distribution than systems constructed with regenerating codes. Moreover, we constructed a new secure storage framework in presence of passive eavesdroppers. Two types of attack were taken into consideration. First one, when the eavesdropper has an access in the data stored on nodes in the the cloud storage system and second one, when the eavesdropper is observing the data downloaded during the repair process. We proved that when the eavesdropper has an access in the stored data, the secure file size in this new storage framework is greater than the secure file size in the DSS modeled with MBR codes, for $\beta = 1$, $\alpha = k$ in presence of $l = l_1$ $(l < k)$ eavesdroppers; and in an eavesdropper model with both types of attacks our storage framework supports greater secure file size than the secure file size supported with the MSR codes, for $\beta = 1$ and $d = 2k - 1$ in presence of $l_1 + l_2 < k$ eavesdroppers.